# Kinetic ionization and recombination coefficients in the dense semiclassical plasmas on the basis of the effective interaction potential

**E O Shalenov, M M Seisembayeva, K N Dzhumagulova and T S Ramazanov**

IETP, Department of Physics and Technology, al–Farabi KazNU, al–Farabi 71, 050040 Almaty, Kazakhstan

E-mail: shalenov.erik@mail.ru

**Abstract**. In this paper, the ionization and recombination coefficients of dense semiclassical hydrogen plasma on the basis of the effective interaction potential have been investigated. For this goal the Bohr–Lindhard method and method phase function have been applied to obtain the electron capture and ionization cross sections. The electron capture cross section has been calculated in the framework of the perturbation theory. The effective interaction potential, which takes into account the screening effects at large distances and quantum diffraction effects at short distances, was used. The results of the investigation show the behaviour of the calculated kinetic coefficients with a change in the plasma parameters: the ionization coefficient decreases with increasing density and (or) coupling parameter while the recombination coefficient increases.

## 1. Introduction

To solve the actual problems of thermonuclear controlled fusion (TCF) with inertial confinement, as well as to study the processes taking place in astrophysical objects (white dwarfs, the Sun, the bowels of giant planets, etc.), reliable data on the physical characteristics of the nonideal semiclassical plasma arising on Earth and in the Cosmos in many processes associated with the heating and compression of matter is needed. Dense semiclassical plasma is observed, for example, when the target substance is compressed by a high–power laser radiation in nuclear fusion, in nuclear explosions, at supersonic motion of bodies in dense layers of planetary atmospheres, at impact of high–intensity energy fluxes on the surface of various materials. Femtosecond laser pulses with high intensities are available now to produce hot dense plasmas in the laboratory. The development in the fields of laser produced plasmas [1] and heavy ion beam plasma [2] interactions led to a growing interest in the kinetics of ionization [3] and recombination [3–4] and also in other processes in the dense semiclassical plasmas [5–11].

Recombination can take place at the collision of an electron with an ion if the latter capture the electron. The process of electron capture by an atom has been investigated in many studies [12–13]. A neutral hydrogen atom can be converted to a negative hydrogen ion because of the polarization capture of the electron. In [13] the electron capture cross section was theoretically considered, and a method for finding the capture radius based on perturbation theory was proposed. To find the radius, time, and probability of electron capture, the Bohr–Lindhard method [12–13] was applied. In Ref. [13] for calculation of the electron capture cross section the applying of the calculated electron trajectories near target particle instead of the linear trajectories, which are used in the perturbation theory [12], has been made. In the present work we have used this approach to calculate the cross section of the electron capture by the hydrogen ion (proton) on the basis of the numerical simulation of the equations of electron motion near proton. The effective potential of electron–ion interactions, taking into account the effects of static screening and diffraction [5–14] has been used for this goal.

## 2. Effective interaction potential

In dense plasma at numerical densities $n = 10^{20} \div 10^{24} \, cm^{-3}$, quantum-mechanical effects begin to play a pivotal role, since the average interparticle distance decreases and becomes of the order of the de Broglie thermal wavelength of the interacted particles. In this case, if the plasma is not degenerate yet, then it can be considered as semiclassical and the effects of degeneration can be neglected while effects of diffraction or symmetry are taken into consideration. The effective potential of electron–ion interactions, taking into accounts the effects of static screening and diffraction [5–14], can be written as

$$\Phi_{ei}(r) = -Ze^2 \left(rC_{ei}\right)^{-1} \left(e^{-B_{ei}r} - e^{-A_{ei}r}\right), \tag{1}$$

here $A_{ei}^2 = (1+C_{ei})/(2\lambda_{ei}^2)$, $B_{ei}^2 = (1-C_{ei})/(2\lambda_{ei}^2)$, $C_{ei}^2 = (1-4\lambda_{ei}^2/r_D^2)$. $\lambda_{ei} = h/\sqrt{2\pi\mu_{ei}k_BT} \approx \lambda_e$ is the de Broglie thermal wavelength; $\mu_{ei} = m_e m_i/(m_e + m_i)$ is the reduced mass of the ion and the electron; $r_D = \left(k_BT/(4\pi e^2 n_e)\right)^{1/2}$ is the Debye length; $n_e$ is the numerical density of electrons; $T$ is the plasma temperature; $k_B$ is the Boltzmann constant, and $Z$ is the charge number of the ion. In this work the following dimensionless parameters were used: $\Gamma = Ze^2/a k_BT$ is the coupling parameter, the average distance between particles is $a = (3/4\pi n)^{1/3}$, $n$ is the density of the charged particles. The density (Brückner) parameter $r_s = a/a_B$ ($a_B = \hbar^2/m_e e^2$ is the Bohr radius, $\hbar$ is the reduced Plank constant). The effective potential (1) is presented in figure 1 as function of the distance between particles for different values of the parameter $r_s$. It is seen that the effective potential has finite values at the distances close to zero. At high distances, the effective potential tends to the Debye potential, which takes into account the effect of screening.

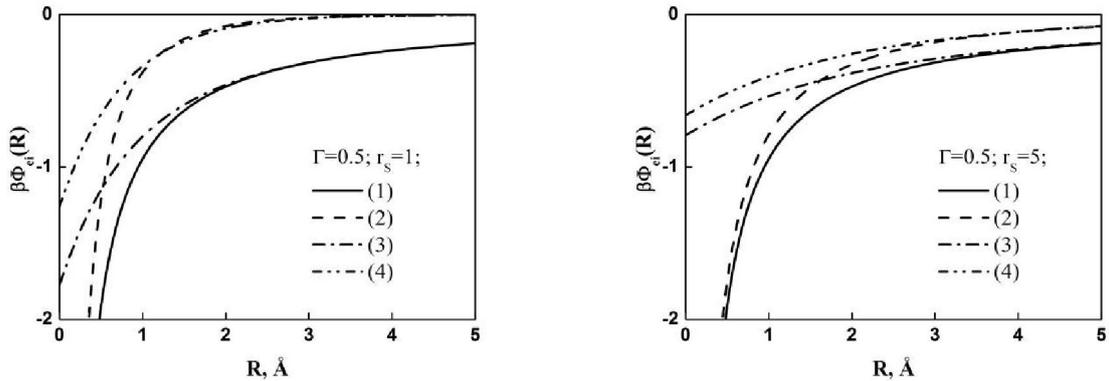

**Figure 1.** Potentials of the electron–ion interaction for different density parameters. 1 – Coulomb potential; 2 – Debye potential; 3 – Deutsch potential; 4 – effective potential (1).

## 3. Theory and results

The Saha equation allows us to determine the number of particles of different types per unit volume for the case when the plasma is in a state of thermodynamic equilibrium. In the general case, the composition of the plasma is determined on the basis of the so-called ionization kinetics equations. These equations describe the rate of change in the number of particles (growth or decrease) in a certain state (free, bound, etc.), due to various reactions. So the rate of growth of the number of electrons per unit volume as a result of ionization of an atom by electron impact is described by equation

$$\frac{dn_e}{dt} = n_e n_a \alpha, \qquad (2)$$

where $\alpha = \langle \sigma_i v_e \rangle$, the brackets mean averaging over the electron distribution. If there are many electrons, the reverse process of triple recombination begins to play a role:

$$H^+ + e + e \to H + e. \qquad (3)$$

The number of collisions leading to recombination is $\beta n_e^2 n_i$, where $\beta = V_* \langle \sigma_* v_e \rangle$ is the rate constant of recombination, which depends on the electron temperature. Thus, the rate of change in the number of electrons per unit volume is

$$\frac{dn_e}{dt} = n_e n_a \alpha - \beta n_e^2 n_i. \qquad (4)$$

If the ionization cross section is known, the ionization coefficient can be calculated as follows [15–16]:

$$\alpha = \int_{p_i}^{\infty} \sigma_{ion}(p) f(p) p \, dp, \qquad (5)$$

where $\sigma_{ion}(p)$ is the ionization cross section, $p$ is the momentum of the incident particle, $p_i = \sqrt{2m_e I}$ corresponds to the ionization energy, $f(p) = 4\pi p^2 \exp\left[-p^2/(2m_e k_B T)\right] / (2\pi m_e k_B T)^{3/2}$ is the Maxwell momentum distribution function, then we can rewrite (5) as:

$$\alpha = g \int_{k_i}^{\infty} dk \, k^3 \, \sigma_{ion}(k) \exp\left[-\hbar^2 k^2 / 2m_e k_B T\right], \qquad (6)$$

where $g = 4\pi \hbar^4 / \left(2\pi m_e^{5/3} k_B T\right)^{3/2}$. The recombination coefficient is written here [16–17], where we used electron capture cross section $\sigma_{cap}(p)$ as the recombination cross section:

$$\beta = \lambda_e^3 \exp[I/k_B T] \int_0^{\infty} \sigma_{cap}(p) f(p) p \, dp, \qquad (7)$$

where $s_{cap}(p)$ is the cross section for electron capture by an ion.

To calculate the recombination coefficient, it is necessary to know the electron capture cross section of an ion. To determine it, we used the Bohr-Linhard method, which was described in Ref. [13]. Figure 2 shows plot of the ionization coefficient versus the coupling parameter calculated for different values of the density parameter. We see that as the coupling parameter increases, the ionization coefficient decreases. It can be seen from the figure that with a decrease in the coupling parameter, the ionization coefficient increases. The figure of the dependence of the ionization coefficient on temperature, calculated by eq.(6) with ionization cross section [18–19] calculated on the basis of the phase functions, is presented in figure 3. As can be seen from this figure, the ionization coefficient on the basis of the effective interaction potential good agreement with other authors.

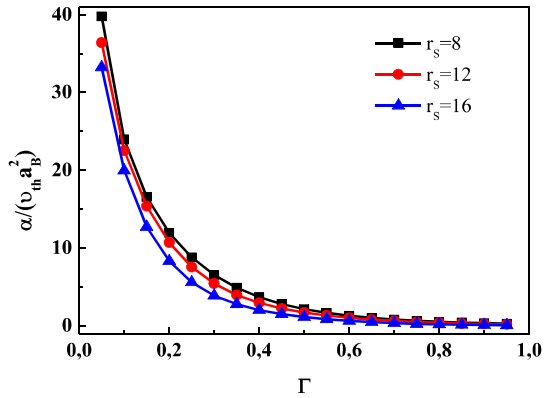 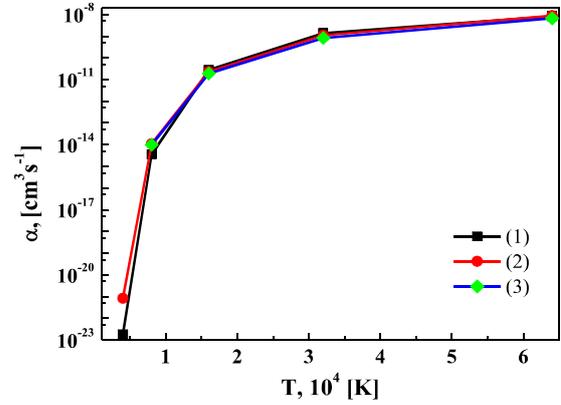

**Figure 2.** Dependence of the ionization coefficient on the coupling parameter for different values of the density parameter.

**Figure 3.** Dependence of the ionization coefficient on temperature. 1 – based on the static potential, 2 – [20], 3 – [21].

Figure 4 shows a comparison of the differential cross sections for the capture of an electron by a hydrogen atom and a proton. Here the cross sections are calculated on the basis of perturbation theory. As can be seen from this figure, the cross section for the capture of an electron by a proton lies higher than the capture cross section by an atom, since the interaction of charged particles is more intense than the charge-atom.

The graph of the dependence of the recombination coefficient on the coupling parameter for different values of the density parameter, calculated by the perturbation theory, is presented in figure 5. It can be seen that as the density decreases (increases the parameter $r_s$), the recombination coefficient climbs, since the screening is weakened (the energy of the interaction of the electron and the positive ion increases).

## 4. Conclusion

Based on the effective interaction potentials, which take into account the screening effect at large distances and the diffraction effect at short distances, the ionization and recombination coefficients of the dense semiclassical plasma have been estimated.

In studying the electron capture process, the motion of an electron was considered on the basis of perturbation theory (rectilinear trajectories). The results of the investigation show the characteristic behavior of the calculated kinetic coefficients with a change in the plasma parameters: the ionization coefficient decreases with increasing density and (or) coupling parameter, but the recombination coefficient increases.

## 5. Acknowledgments

The authors acknowledge support within the Program BR 05236730 of the Ministry of Education and Science of the Republic of Kazakhstan.

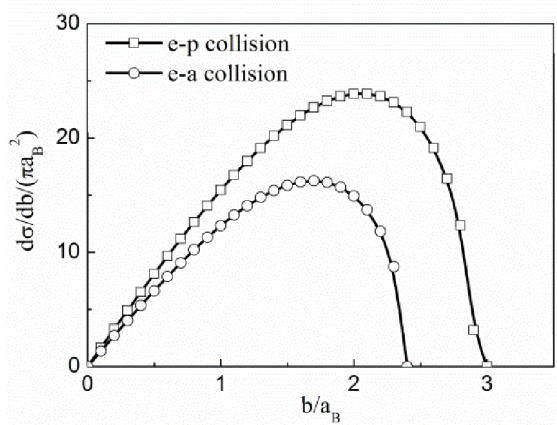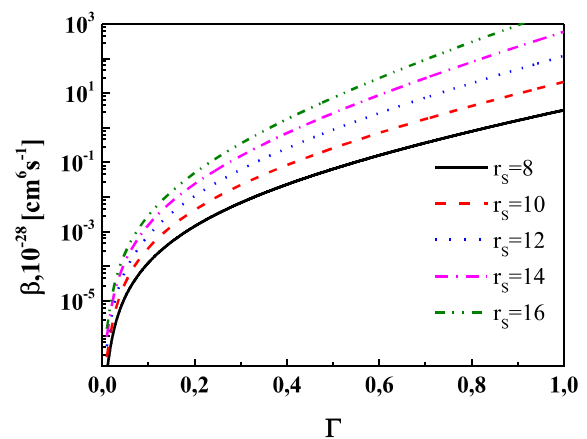

**Figure 4.** Comparison of the differential cross sections for the capture of an electron by a hydrogen atom or a proton, calculated on the basis of perturbation theory for $G = 0.2$, $r_s = 10$.

**Figure 5.** Dependence of the recombination coefficient, calculated on the basis of perturbation theory, on the coupling parameter for various values of the density parameter.